# When Atoms Choose Their Neighbors: Element-Specific Views of Local Chemical Order in High-Entropy Alloys

*David Morris[1], Yonggang Yao[2], & Peng Zhang[\*,1]*

[1]Department of Chemistry, Dalhousie University, Halifax, NS, Canada.

[2]State Key Laboratory of Materials Processing and Die & Mould Technology, School of Materials Science and Engineering, Huazhong University of Science and Technology, Wuhan, China.

ABSTRACT

Local chemical order (LCO) is a key descriptor linking composition, atomic arrangement, and function in high-entropy alloys (HEAs), yet remains difficult to quantify. This Perspective highlights how X-ray absorption spectroscopy (XAS) provides element-specific, quantitative insight into LCO in complex alloys. We outline practical considerations for XAS data collection and fitting, including width of spectra range, multi-temperature analysis, and X-ray absorption edge choice for 5d elements. We then highlight a coordination-number-based framework for LCO analysis and use model HEAs to show how single-atom and near-single-atom motifs naturally emerge as component number increases, bridging HEAs and single-atom alloys. Finally, we identify priorities including standardized protocols, uncertainty quantification, expanded operando and time-resolved XAS, and integration with complementary characterization and modelling.

1. INTRODUCTION: HEAS AND LCO

High-entropy alloys (HEAs) are often described as a class of materials composed of five or more elements in approximately equiatomic ratios.[1–7] A key feature of HEA structures is that, despite their chemical complexity, they often form simple solid-solution phases, in contrast to the heavily segregated multiphase structures typical of conventional multi-element alloys.[8–11] This ability to stabilize disordered solid solutions allows HEAs to overcome many of the constraints associated with conventional alloy design, unlocking a vast, high-dimensional compositional space that in turn enables fine-tuning of the material's properties.[12] The versatility of HEAs has been demonstrated across a wide variety of fields, including structural applications with exceptional strength and stability, energy storage systems with high electrochemical stability and long cycle life, and catalytic processes with remarkable activity and selectivity.[13–22] To fully realize the potential that HEAs have to offer, it is vital to establish a clear understanding of their structure-property relationships, particularly with respect to how atomic-scale arrangements influence macroscopic behaviour.

One key concept for understanding the structure-property relationships of HEAs is local chemical order (LCO). LCO is commonly defined as ordering on the scale of approximately one nearest-neighbor bond length, capturing the tendency of atoms to adopt preferred arrangements rather than remaining in a perfectly random distribution.[23] LCO can be utilized as a quantitative measure of the deviation from a perfectly mixed HEA, in which all atomic pairings are assumed to occur with equal probability.[24] As such, LCO has substantial overlap with the concept of chemical short-range order (SRO), which is commonly defined in alloy theory in terms of deviations of nearest-neighbor pair probabilities from those expected in a random solution.[25] In practice, LCO and SRO are used interchangeably in the HEA literature to describe deviations from random nearest-neighbor correlations.[26] In this Perspective, we will use LCO as the term to describe this phenomenon. LCO includes both local clustering, in which like-element neighbors are enriched relative to random mixing, and chemical ordering, in which unlike-element neighbors are enriched relative to random mixing.[27,28] LCO describes ordering at the nearest-neighbor scale, in contrast to long-range order, which reflects ordering across multiple coordination shells.[25]

LCO plays a significant role in modifying the properties that an HEA possesses. For example, preferential clustering can impact dislocation motion, which in turn can alter the mechanical strength of a material,[29–31] while selective nearest neighbor-interactions can modify the local electronic structure of a catalytically active site, changing the activity of the catalyst.[23] Understanding the LCO an HEA possesses is therefore critical for the rational design of compositions tailored for specific applications. Achieving a detailed characterization of LCO has remained a significant challenge, motivating the development of advanced element-specific characterization tools that can probe order at the atomic scale.[32,33]

Despite its importance, LCO remains difficult to measure and evaluate accurately in HEAs. As LCO manifests at the scale of a single nearest-neighbor bond length, characterization tools with exceptionally high sensitivity are necessary to detect and quantify subtle deviations from random mixing. This makes many of the widely used conventional characterization tools in materials science insufficient. For example, techniques such as X-ray diffraction are useful for collecting information regarding the long-range order of a sample, but they lack the resolution to identify details regarding short-range chemical arrangements. High-resolution imaging methods such as transmission electron microscopy (TEM) have been widely applied to reveal atomic-scale features, but these methods encounter challenges regarding sampling statistics, susceptibility to imaging artifacts, and the difficulty of distinguishing elements of similar atomic number.[33,34] This can therefore result in misleading or inconclusive conclusions being drawn regarding the LCO of an HEA sample. These persistent challenges have made it difficult to establish a clear correlation between the LCO of an HEA sample and the properties it possesses. Addressing these challenges requires characterization approaches that offer both atomic-scale sensitivity and element specificity.[23]

Another challenge that emerges when characterizing HEAs is the sheer volume of data that must be analyzed. While early HEAs contained relatively low numbers of elements, advances in synthesis techniques have dramatically expanded the compositional space to nanoparticle (NP) systems containing up to 21 different elements.[3,35,36] This large compositional space offers exciting opportunities to explore new combinations of properties, but it also introduces significant challenges for characterization. This is especially true when considering LCO, as the number of distinct element-specific bonding interactions, $I$, that must be characterized grows as $I = N^2$ for an HEA with $N$ constituent elements (Figure 1). In other words, each element can, in principle, adopt all $N$ possible nearest-neighbor environments, making it extremely difficult to identify and quantify the contribution of each interaction to the overall properties. Element-specific techniques such as X-ray absorption spectroscopy (XAS) are well-suited for this task, but fully resolving all relevant absorber-neighbor pairs requires collecting and fitting large numbers of spectra, introducing substantial logistical constraints on the characterization process.[24] As advances in HEA synthesis continue to expand the accessible compositional space, this will simultaneously increase the need for sophisticated characterization methodologies.

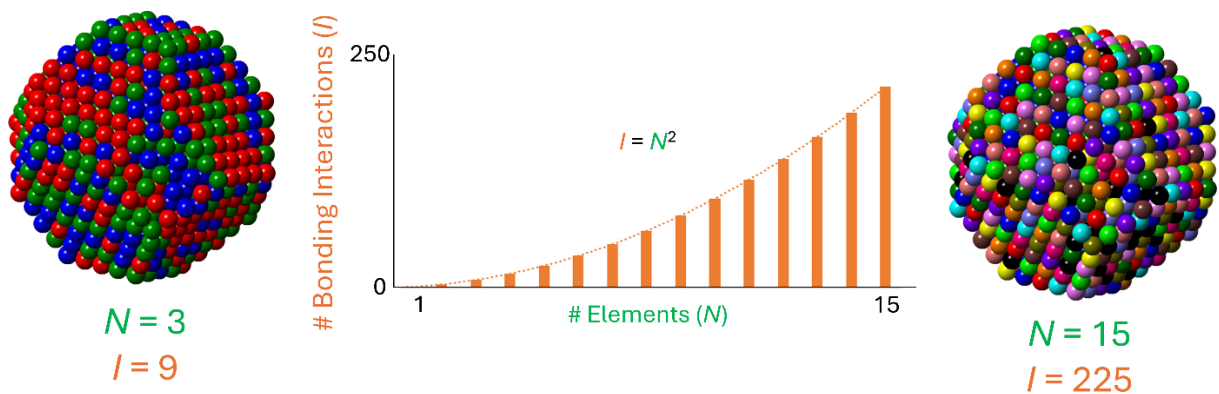

**Figure 1.** Impact of compositional complexity on the challenge of quantifying LCO in HEAs. The number of element-specific bonding interactions, $I$, grows with the square of the number of constituent elements, $N$ ($I = N^2$), dramatically increasing the amount of structural information that must be extracted to fully characterize LCO. Each distinct elemental bonding pair requires independent analysis to determine deviations from random mixing, making characterization progressively more complex as the number of elements increases. This quadratic combinatorial growth underscores the importance of adopting element-specific, quantitative techniques such as XAS, which can isolate and analyze individual bonding interactions even in multi-element systems.

A further challenge in the study of LCO is determining how best to quantify it. Several approaches have been utilized throughout the literature, yet there is currently no universal metric for describing LCO across different systems.[37–39] In many cases, the choice of the exact metric being used can influence whether a material is considered to be "ordered" or "random," making it extremely difficult to compare materials across different studies. Additionally, experimental uncertainties can be large enough that apparent deviations are not statistically significant, raising further challenges for rigorous interpretation. In our recent work, we have sought to address this problem by developing an approach that propagates the uncertainty in XAS-derived coordination numbers (CNs) into the calculation of a $\Delta P$ parameter, allowing for a robust and statistically defensible assessment of LCO in HEAs.[24]

An additional complication in the study of LCO is its inherently dynamic nature. The degree and character of LCO can change under different conditions, for example with variations in temperature, applied stress, and chemical environment.[40,41] This is of particular importance in the context of catalysis, as the local arrangement of atoms at the active sites can change under operating conditions, potentially impacting activity, selectivity, and stability.[42] Despite this, to date the majority of studies of LCO in HEAs have been carried out under static conditions, providing only a single snapshot of LCO for materials that are likely dynamic in nature.[23,31] As a result, our current understanding does not account for how LCO fluctuates and adapts under the conditions that applications will expose HEAs to. Understanding the dynamic nature of LCO will require the development of in situ and operando characterization methods capable of probing LCO as it evolves, an area where element-specific spectroscopy holds particular promise.[43]

Taken together, these considerations highlight both the promise of HEAs and the critical gaps that remain in our understanding of their local structure. While LCO is increasingly recognized as a key factor impacting HEA properties, its measurement, quantification, and evolution under working conditions remain significant challenges. Addressing these issues requires techniques that combine atomic-scale sensitivity with element-specific resolution, enabling reliable insights into the short-range interactions present within complex systems. In this perspective, we begin by presenting a brief survey of various characterization approaches that have been used in the study of HEAs, outlining the strengths and limitations that each has with respect to the analysis of LCO. We then discuss emerging opportunities in this field, particularly focusing on how element-specific spectroscopy can play a central role in advancing the study of LCO in HEAs.

2. PROBING LCO IN HEAS: CAPABILITIES AND LIMITATIONS

Accurate characterization of the structure of HEAs is vital for developing a clear understanding of the nature of LCO and the role it plays in the properties of the material. A wide range of characterization tools have been applied to serve this purpose, each possessing its own strengths and weaknesses (Table 1). These tools include diffraction methods, which excel at analyzing long-range order, imaging approaches, which allow for direct visual analysis at the atomic scale, and spectroscopy tools, which yield information about the local bonding environments. In this section, we provide a brief survey of each of these commonly used tools, outlining their respective strengths and limitations, before turning to the opportunities offered by element-specific spectroscopy.

**Table 1**. Summary of characterization tools utilized to probe LCO in HEAs.

| Technique | Primary Sensitivity | Spatial Resolution | Element Specificity | Sampling Volume |
|---|---|---|---|---|
| **XRD** | Long-range periodic order | Sub-Å | Limited (scattering contrast) | Very high (ensemble) |
| **PDF** | Real-space pair distances | Sub-Å | No | High (ensemble) |
| **TEM/HAADF-STEM** | Local contrast | Å | Limited | Low |
| **STEM-EDS/EELS** | Elemental maps (local compositional fluctuations) | Å | Yes | Low |
| **APT** | 3D atomic-scale composition | nm | Yes (mass-to-charge) | Moderate |
| **AET** | 3D atomic positions | Sub-Å | Limited (for similar Z) | Very low |
| **XPS** | Surface composition | N/A | Yes | Surface only (few nm depth) |
| **XAS** | Local structure | Sub-Å | Yes (edge-selective) | Very high (ensemble) |

Diffraction methods, particularly X-ray diffraction (XRD), have been widely employed in the characterization of HEAs.[5,6] Conventional XRD allows for the precise determination of lattice parameters, the identification of phase segregation through the presence of secondary peaks, and the detection of long-range strain and ordering. This is key information when confirming whether or not an HEA has a single-phase structure, as is often expected given the synthesis conditions. While XRD is useful for identifying long-range ordering behavior, it lacks the sensitivity to characterize LCO.[23] XRD also struggles with the characterization of samples that are amorphous
8

or highly-disordered, and it is difficult to distinguish elements of similar atomic number due to similarities in their scattering factors.[44] To help address these limitations, many studies have utilized XRD in conjugation with pair distribution function (PDF) analysis.[45,46] PDF analysis transforms scattering data into a real-space distribution of atomic pair distances, which allows for the determination of local structure in even highly disordered materials.[47] Experimental PDFs can be compared to simulated PDFs representing a perfectly random alloy to quantify the LCO present in an experimental sample.[48–50] While this is a powerful approach capable of quantifying LCO, PDFs are ensemble-averaged and not element-specific, which limits the ability to use this methodology in complex HEAs containing many overlapping atomic pair correlations.

TEM and scanning transmission electron microscopy (STEM) are powerful techniques that allow for the direct visualization of the structure of HEAs through real-space imaging. When combined with high-angle annular dark-field (HAADF) imaging and spectroscopic methods such as energy-dispersive X-ray spectroscopy (EDS) or electron energy-loss spectroscopy (EELS), it is possible to generate element-specific maps for the sample at near-atomic resolution, even in samples containing elements of similar atomic number.[51–55] Through the real-space data collected using this approach, local compositional variations consistent with LCO can be directly visualized. However, atomic-resolution images require highly specialized instrumentation and careful sample preparation, making this a relatively inaccessible technique. Moreover, the nature of imaging techniques means that only a small fraction of the overall sample is being measured, resulting in limited sampling statistics and potential selection bias, which in turn leads to uncertainty in conclusions regarding LCO in the sample.[56,57] This makes it difficult to accurately quantify the overall LCO of the sample using imaging approaches.

To overcome the limited sampling volume of two-dimensional imaging techniques, three-dimensional compositional mapping techniques have emerged as powerful tools for the characterization of HEAs. Atom probe tomography (APT) offers near-atomic resolution by field-evaporating atoms from a needle-shaped specimen and reconstructing their positions in three dimensions. This technique allows for the determination of LCO with a much larger dataset compared to conventional imaging techniques.[58,59] However, APT reconstructions are susceptible to field-evaporation artifacts, preferential loss of certain species, and overlap of mass-to-charge peaks, which can complicate quantitative data analysis.[60,61] In recent years, atomic electron tomography (AET) has emerged as a powerful tool capable of reconstructing three-dimensional atomic configurations with sub-angstrom precision, enabling the atomic structure of individual NPs to be resolved.[2,33,62] This allows for the direct evaluation of LCO at the atomic level beyond what is possible with conventional techniques. AET does have some limitations, for example it requires specialized equipment and is computationally demanding. This leads to similar selection bias issues, as obtaining a map for even a single NP within the sample is very demanding. AET is also not fully element-specific, as it is incapable of distinguishing elements of similar atomic number.[63] This is particularly problematic in HEAs of increasing complexity, in which there are likely to be many elements of similar atomic number. Furthermore, while APT and AET can provide geometric local environments from reconstructed 3D atomic configurations, they do not directly probe electronic structure. These limitations motivate the complementary use of element-specific spectroscopic techniques.

Surface-sensitive spectroscopic tools, such as X-ray photoelectron spectroscopy (XPS), have been widely utilized to provide important insights into the structure and properties of HEAs.[64–66] XPS is capable of identifying the elemental composition of a sample's surface, probing the oxidation states of each element, and can detect surface segregation or preferential enrichment of specific elements. While this information is useful for general characterization purposes, XPS is not well suited for the quantification of the LCO in a sample. Due to its limited sampling depth, information obtained from XPS is relevant only to the surface of the sample, and bulk information can therefore only be accessed for sufficiently small NP samples.[67] Moreover, distinguishing between elements with overlapping core-level binding energies, such as Co and Ni, is challenging.[68] Finally, XPS provides no direct information regarding nearest-neighbor interactions or the local bonding geometry.[69] These limitations motivate the use of element-specific spectroscopic techniques that are sensitive to atomic coordination, such as X-ray absorption spectroscopy (XAS).[70]

XAS offers a powerful approach for probing LCO in HEAs. Unlike diffraction or imaging techniques, XAS is inherently element-specific, allowing for the local chemical environment around each constituent element to be isolated and quantified.[12,71–76] Its flexible sample preparation requirements allow it to be applied to powders, thin films, bulk samples, colloidal NPs and more. Operando measurement capabilities enable the direct observation of structural changes under various conditions, thereby enabling the evaluation of LCO under dynamic conditions.[77–80] By analyzing the extended X-ray absorption fine structure (EXAFS) region, CNs and interatomic distances can be calculated, providing quantitative insights into LCO.[81–83] The element-specificity and focus on the local structure allow LCO to be decoupled from long-range order, which is especially valuable in disordered systems where diffraction offers little contrast.

Despite these advantages, there are still some limitations associated with using XAS to evaluate LCO in HEAs. XAS represents an ensemble average over the entire sample, which does not allow for spatially resolved assessment of LCO as in techniques such as AET. While XAS is element-specific in terms of collecting separate data for each element, the scattering contributions from elements of similar atomic number are difficult to distinguish, and the limited number of independent parameters available in an EXAFS fit often necessitates grouping multiple bonding interactions into shared fitting paths in multi-element systems.[12] Absorption edge overlap can also become an issue, particularly in samples containing many 5d transition metals, limiting the amount of data available for use in the fitting procedure.[24]

When viewed together, these characterization tools reveal a set of complementary but still incomplete insights into LCO in HEAs. The relative usefulness of each technique largely depends on the exact system being studied. For alloys with a relatively small number of constituent elements, tools such as AET and PDF can provide detailed local structural information with manageable complexity. However, as the number of constituent elements, $N$, in the HEA increases, the number of unique atomic pairings grows as $N^2$, making it increasingly difficult for diffraction and imaging techniques to resolve overlapping contributions. In this context, the element-specific nature of XAS becomes a major advantage, enabling the environment of each constituent element to be interrogated independently. This scalability makes XAS particularly well-suited for next-generation HEAs, where compositional complexity is high and disentangling contributions from multiple species is critical to building a complete understanding of LCO.

Recent studies have increasingly applied XAS to investigate LCO in HEAs through EXAFS fitting of each constituent element.[12,24,35,48,74,84–88] However, efforts to compare LCO across systems have been hindered by the lack of robust, standardized approaches for interpreting these findings.[89] In our recent work,[24] we introduced an approach that compares experimental values to those that would be expected from a perfectly-mixed random alloy, allowing for rigorous quantification of LCO for each absorber-neighbor pair in the sample (Figure 2). Briefly, for each absorber-neighbor pair (A-B), EXAFS fitting provides the first-shell partial coordination number, $CN_{AB}$, and the total first-shell coordination number around the absorber, $CN_{AM}$. Their ratio gives the ensemble-averaged nearest-neighbor occupancy probability, $P_{AB}$, which is the average fractional occupancy of B in the first coordination shell of A. We define the random-alloy reference probability from the bulk composition as $C_B$, and compute $\Delta P$ as the deviation of $P_{AB}$ from $C_B$ (see Supporting Information for additional detail). Positive $\Delta P$ indicates enrichment of B around A relative to random mixing (chemical ordering), whereas negative $\Delta P$ indicates depletion (segregation tendency). For cross-field comparison, $\Delta P$ is conceptually related to the Warren-Cowley SRO parameter ($\alpha_{AB}$) as both compare nearest-neighbor pair statistics to a random-alloy baseline; however, $\Delta P$ remains straightforward to interpret and compare across multicomponent and dilute-component alloys, where $\alpha_{AB}$ can become unintuitive (see Supporting Information for additional detail).

Because $\Delta P$ is derived from first-shell EXAFS coordination numbers, it is inherently a nearest-neighbor metric. As XAS is an ensemble-averaged technique, spectra from multiphase samples reflect a weighted average of all local environments present in the sample. In cases where ordered and disordered phases (or distinct phases) have very similar first-shell coordination geometries and compositions, first-shell EXAFS fitting may yield similar CN (and therefore $\Delta P$) values. Resolving such cases may require multi-shell EXAFS fitting, and the use of complementary characterization techniques (e.g., XRD, PDF, microscopy) to distinguish long-range order and phase fractions.

Crucially, this approach incorporates full propagation of the uncertainties from the EXAFS fitting, allowing statistically significant LCO to be distinguished from results that are consistent with random mixing. To aid interpretation, we complemented the XAS-based analysis with atomistic models derived from computational methods, which help visualize the local configurations indicated by the EXAFS fitting results. This methodology addresses a key gap in the field by improving reproducibility and enabling rigorous comparison across studies. Together, these developments position XAS as a particularly promising approach for routine, quantitative assessment of LCO, helping to set the stage for future efforts to establish it as a standard tool in HEA research.

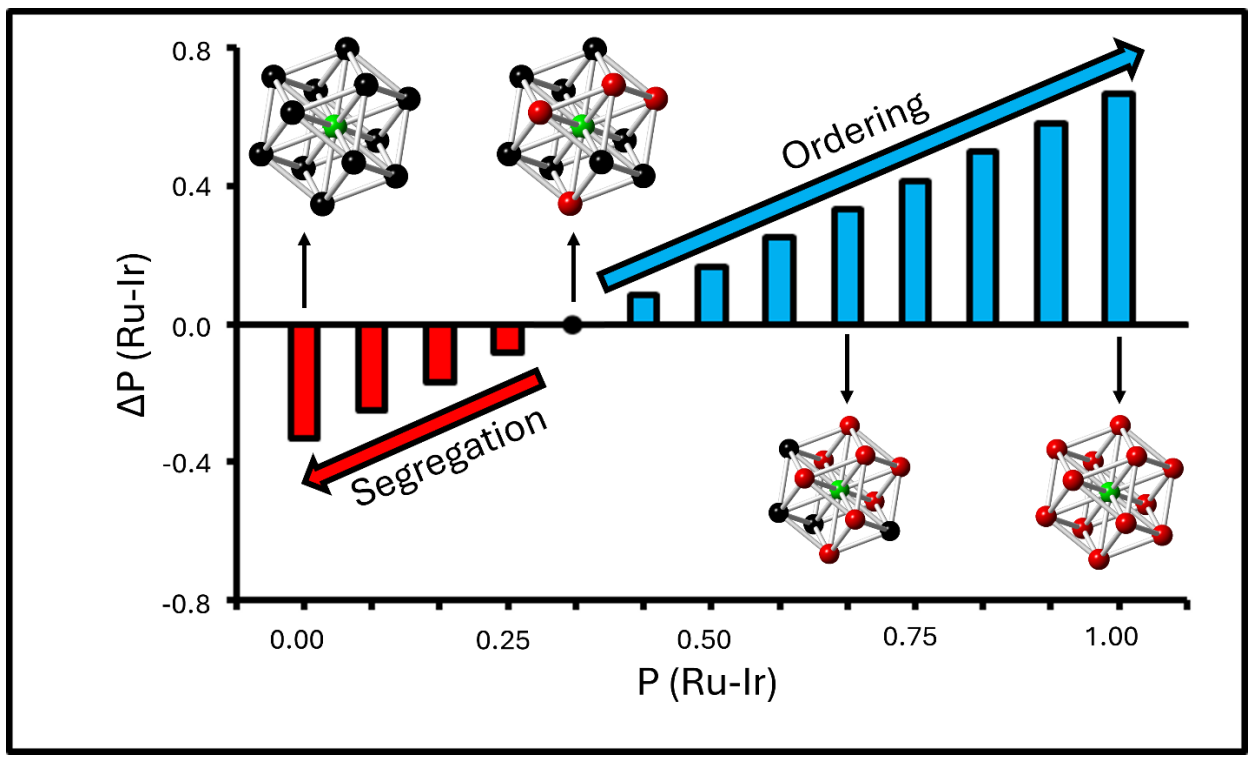

**Figure 2.** Conceptual visualization of the ΔP metric for quantifying LCO. Bars show ΔP(Ru-Ir) as a function of P(Ru-Ir), the fraction of Ir atoms in the first coordination shell around a central Ru atom, for an equiatomic ternary system containing Ru (green), Ir (red), and a generic third element (black). For a perfectly mixed alloy, P(Ru-Ir) = 0.33 and ΔP = 0. Enrichment of Ir in the first coordination shell corresponds to ordering and yields positive ΔP values (blue bars), while depletion of Ir indicates segregation and produces negative ΔP values (red bars). Representative 13-atom first-shell configurations are shown for selected P(Ru-Ir) values, illustrating the progression from segregation to ordering. CNs derived from EXAFS fitting enable calculation of ensemble-averaged ΔP values for each bonding interaction in the sample, together with their associated uncertainties. Interpreting these values provides quantitative evidence for ordering, random mixing, or clustering, enabling a statistically robust assessment of LCO across varying samples.

3. ADVANCING ELEMENT-SPECIFIC LCO ANALYSIS IN HEAS

The progress made in characterizing LCO in HEAs in recent years has established its importance as a key structural parameter influencing material properties. Despite this, our current understanding remains incomplete, limited by the difficulty of reliably measuring LCO, the absence of universally standardized quantification methods, and the scarcity of studies that capture its evolution under dynamic conditions. These challenges result in significant barriers to establishing clear structure-property relationships, which in turn inhibits the rational design of next-generation HEAs. In this section, we outline opportunities to overcome these barriers through the advancements of both experimental and analytical approaches. In particular, we focus on how XAS has the potential to serve as a routine and standardized technique for probing LCO in HEAs, and discuss how its integration with operando measurements, computational modeling, and data-driven analysis can help to advance this field.

A major opportunity for advancing the study of LCO in HEAs lies in establishing standardized, reproducible methods for its quantification. Currently, a range of metrics has been employed across the literature,[2,23,24,90–92] complicating comparisons and limiting the ability to extract generalizable design principles. Recent efforts, such as our own methodology, demonstrate that rigorous and statistically meaningful quantification is achievable. The next step is to move beyond individual studies towards community-wide adoption of best practices. This could be achieved by developing open-source analysis pipelines for EXAFS fitting, standardizing the reporting of CNs alongside their respective uncertainties, and the creation of databases of LCO metrics for various HEA samples. These efforts will improve reproducibility, facilitate meta-analyses, and allow greater confidence when establishing structure-property relationships.

A particularly exciting opportunity for advancing the study of LCO is to move beyond static characterization and measure its dynamic behavior under realistic conditions. Since LCO can evolve with temperature, chemical environment, or mechanical stress, static ex situ measurements risk missing the transient configurations that may play a key role in controlling material properties. Operando XAS is an ideal solution to probe these dynamics and has the potential to monitor local coordination and LCO-related signatures during catalytic reactions to directly correlate changes in local structure with activity, selectivity, and deactivation pathways. Representative examples have already demonstrated this capability across both processing- and application-relevant conditions, including annealing-dependent changes in element-specific local coordination environments,[93] formation pathways during low-temperature synthesis,[94] and active-site dynamics in HEA electrocatalysts operating under working potentials.[95]

Continuing to extend this approach across various synthesis procedures will allow researchers to trace the formation mechanism of LCO, revealing key insights that can be applied to the fine-tuning of material properties. Although applying operando XAS to extremely fast synthetic processes such as carbothermal shock synthesis presents substantial experimental challenges with current instrumentation, given the ultrafast heating/cooling timescales involved, more conventional synthesis processes are accessible for these types of measurements.[96] Time-resolved XAS approaches, such as quick-EXAFS, can reach millisecond-timescale acquisition in specialized setups, but extending these capabilities to carbothermal shock synthesis-like heating rates and environments remains challenging.[97] Beyond strictly analyzing structural changes arising from LCO, correlating LCO with changes in the local electronic structure is critical for applications such as catalysis. Emerging high-energy-resolution techniques such as high-energy-resolution fluorescence-detected X-ray absorption near-edge structure (HERFD-XANES) offer the potential to detect subtle shifts in unoccupied electronic states as LCO evolves.[98] Overall, operando XAS has the potential to transform LCO analysis from a static descriptor to a dynamic tool for guiding the rational design of HEAs for diverse applications.

No single characterization technique is capable of fully capturing the complexity of LCO in HEAs, making the integration of complementary techniques essential. XAS offers element-specific, quantitative coordination information but cannot resolve atomistic configurations on its own, while imaging methods can directly visualize atomic arrangements but are limited in statistical representativeness. Computational modelling offers a powerful way to bridge these gaps by generating atomistic configurations that are consistent with experimental data, enabling a more complete interpretation of LCO. Combining these techniques can transform LCO analysis from a set of isolated measurements to a coherent structural model. XAS provides quantitative CNs, imaging offers spatial context, and computational modeling offers three-dimensional atomic configurations and energetics. Going forward, machine learning and automated fitting workflows will help to accelerate this process by allowing large volumes of experimental data to be rapidly analyzed and compared. These multimodal strategies not only overcome the limitations of individual techniques but also provide the detailed insights necessary to establish structure-property relationships and guide the rational design of HEAs.

Reliable determination of CNs is essential for quantifying LCO through an XAS-focused approach, making the quality of EXAFS data collection and fitting a critical consideration. High-quality data over a long k-range increases the number of independent data points available for fitting and improves the precision of calculated CNs, in turn strengthening conclusions drawn regarding LCO.[99] Collecting high-quality data often requires averaging many individual scans to improve signal-to-noise ratios, or collecting spectra at multiple temperatures to disentangle static and thermal disorder. Multi-temperature datasets also offer the ability to conduct simultaneous EXAFS fitting, allowing parameters to be constrained consistently across temperatures, effectively increasing the amount of useable information and yielding tighter constraints on fitted parameters.[82,100] These strategies come with the tradeoff of increased beamtime demands, underscoring the importance of continued advancements in beamline optics, detectors, and automation to allow for faster measurements. Emerging techniques such as HERFD can further enhance the quality of the EXAFS data by reducing core-hole broadening.[98] Machine-learning approaches that learn the mapping between EXAFS spectra and local structural descriptors are now being developed for catalysts and disordered materials, and have already been shown to accelerate EXAFS analysis and deconvolute mixed spectra. Extending such methods to multi-element HEAs offers a promising route to disentangle overlapping scattering contributions from elements of similar atomic number and thereby resolve more subtle LCO signatures in these compositionally complex systems.[101–104]

While longer k-ranges, multi-temperature analysis, advanced detection modes, and automated workflows can improve precision, robust CN-based LCO metrics also require careful control of effects that can influence EXAFS amplitudes and therefore CN values. In EXAFS, CN parameters are heavily correlated with amplitude-determining parameters such as $S_0^2$ and $\sigma^2$, meaning that fitting methods that introduce constraints or correlations on some of these parameters can unintentionally lead to bias in others. Constraints or correlations applied across datasets (e.g., across temperatures, edges, or elements) can be powerful for improving fit stability, but they should be applied cautiously and justified explicitly, as differences in absorber identity, edge choice, disorder, or inelastic losses can break the assumption that shared parameters remain equivalent.[105] In XAS fluorescence measurements, self-absorption and thick-sample effects can suppress EXAFS amplitudes in geometric- and composition-dependent manners, impacting CN values if uncorrected. This is particularly relevant in multi-element systems where matrix absorption varies across edges.[106] Finally, because ΔP is derived from ratios of partial CN values, any edge- or element-dependent amplitude bias can potentially mimic preferential enrichment or depletion, leading to false positives of apparent LCO signals. This can arise from background subtraction/normalization choices, inconsistent fitting windows or k-weighting, over-grouping of scattering paths for elements of similar atomic number, unresolved multiphase contributions, or unrecognized changes in oxidation state and coordination that alter disorder and backscattering amplitude.[107] Accordingly, routine CN-based LCO analysis should always involve the transparent reporting of fitting constraints, sensitivity checks to reasonable fitting variants, and complementary structure- or phase-sensitive measurements where ambiguity is possible.

Absorption edge selection is a critical but often underappreciated factor in the EXAFS measurements of HEAs, particularly those containing 5d transition metals. The $L_3$-edge represents electronic transitions from the $2p_{3/2}$ to 5d orbitals, which offers strong absorption due to the large density of unoccupied d-states immediately above the Fermi level. The $L_2$-edge probes $2p_{1/2}$ to 5d transitions, but generally exhibits a weaker absorption step due to spin-orbit splitting and lower transition probability.[108] The $L_1$-edge, which arises from 2s to 6p transitions, is usually less informative and more challenging for quantitative EXAFS because the 2s core-hole has a large natural width and the associated EXAFS oscillations are relatively weak, reducing the practical data quality.[109] As a result, the $L_3$-edge has emerged as the most common choice when collecting EXAFS data for these elements. In HEAs containing many elements of similar atomic number, the relatively small separations between L-edges can lead to edge overlap and limited continuous energy windows free of other edges or fluorescence features (Figure 3). This constrains the usable k-range for EXAFS fitting and therefore reduces the precision with which LCO can be evaluated. Where beamline energy range and sample composition permit, careful selection of alternative edges that offer larger energy gaps could provide an extended k-range that, despite a reduced absorption cross-section, may ultimately yield more precise CNs for LCO analysis.

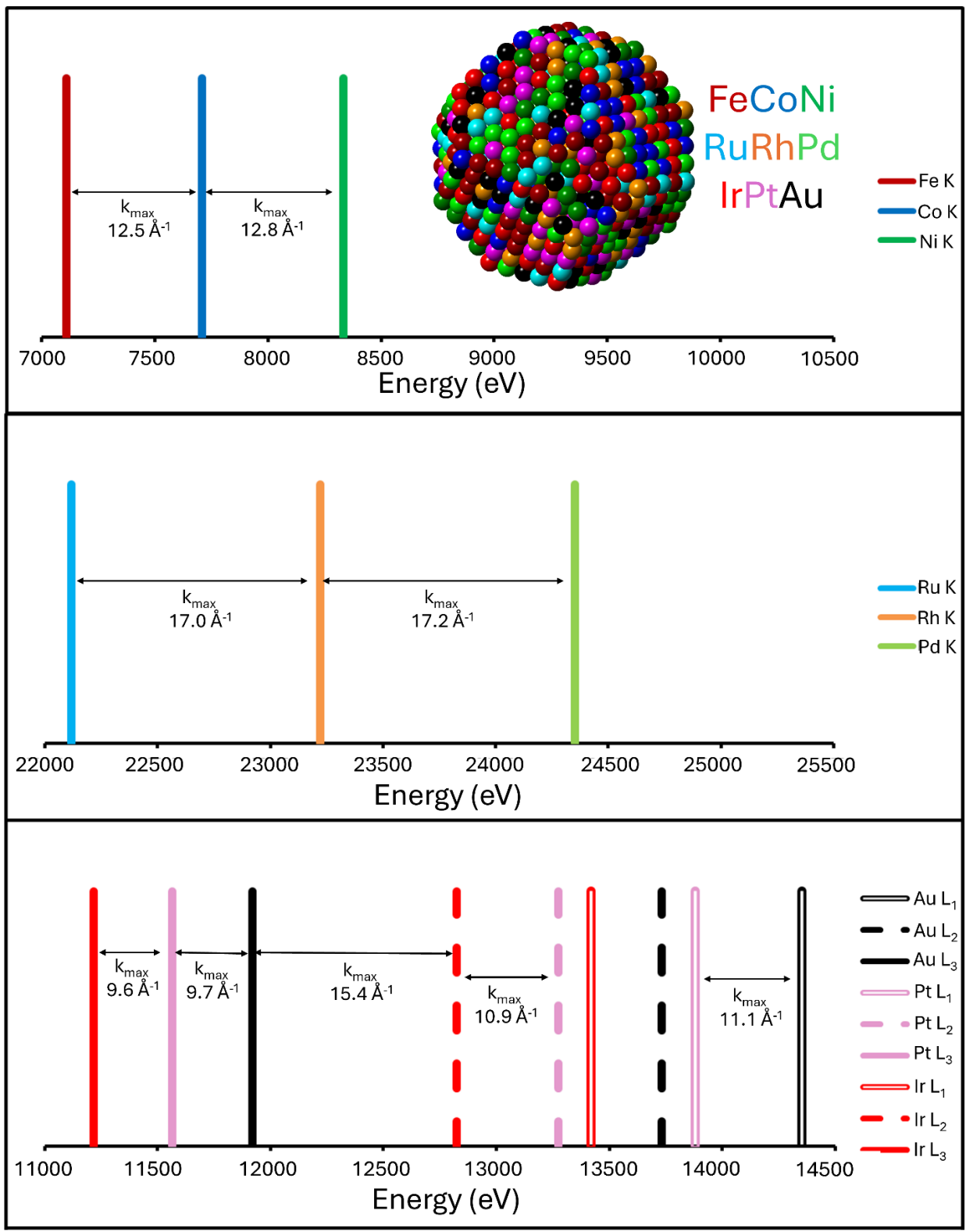

**Figure 3.** Energy separation and maximum usable EXAFS k-range in a hypothetical 9-element HEA. Solid vertical lines represent the conventional choice of K-edges (3d, 4d) and $L_3$-edges (5d). For the 5d elements, values corresponding to the $L_2$ (dashed), and $L_1$ (hollow) absorption edges are also presented. Horizontal arrows indicate the maximum usable k-range ($k_{max}$) for each element before the photoelectron energy reaches the next higher absorption edge in the alloy, beyond which the EXAFS signal is compromised. The tight clustering of $L_3$-edges illustrates the challenge of edge overlap in 5d-rich HEAs, whereas $L_2$ and $L_1$ edges potentially provide wider energy separation and, in turn, a longer usable $k_{max}$. Leveraging these edges offers an opportunity to improve EXAFS resolution and CN precision in some cases, despite the lower absorption cross-sections available at the $L_2$ and $L_1$ edges.

In theory, exciting deeper core levels such as the K-edge, which produces higher-energy photoelectrons with a longer inelastic mean free path, can enhance sensitivity to higher coordination shells in EXAFS.[81] However, the K-edges of 5d transition metals lie at very high energies, accessible only at a small number of specialized synchrotron beamlines.[110] While currently an extremely niche area of study, Pt and Au K-edge XAS has been demonstrated and used to characterize Au-Pt core-shell NPs.[111,112] This proof of concept could be explored further for HEAs containing multiple 5d transition metals. The main obstacle at this stage is that very few synchrotron facilities are equipped for XAS measurements at these high energy ranges, reflecting the limited historical demand and technical challenges for more conventional systems. It is conceivable that as the study of complex HEAs progresses, there will be an increased demand for these specialized 5d K-edge measurements, which may in turn motivate the development of additional high-energy beamlines and make such measurements more accessible. This offers a long-term solution to the issue of absorption-edge overlap, in contrast to the short-term solution of switching to alternate L-edge measurements where appropriate.

In nanoscale HEAs, LCO can be inherently size-dependent due to the large proportion of atoms located at surfaces or interfaces, where coordination numbers and segregation/ordering driving forces differ from those in the bulk. As particle size decreases, these environments dominate an increasing fraction of the material.[113] Consequently, LCO descriptors derived from bulk-averaged measurements should be interpreted with explicit consideration of whether they predominantly reflect bulk-like coordination shells or surface-dominated environments. A practical implication is that nanoscale HEAs offer additional approaches to tune LCO through synthesis and post-synthetic processing. Feasible strategies include kinetic trapping through non-equilibrium synthesis routes, controlled annealing to relax towards preferred nearest-neighbor correlations, and surface/interface engineering to manipulate local environments.[114]

An exciting future direction in this field is to explore the conceptual bridge between HEAs and single-atom alloys (SAAs). When HEAs contain a large number of elements, the atomic fraction of each becomes very low, leading to a high probability of the formation of SAA-like motifs (Figure 4). This is especially true on the surface, where the CNs are lower than in the bulk. Here, we define a "single-atom" (SA) site as an atom that has no like-element neighbors in its first coordination shell. We define "near-single-atom" (near-SA) sites as atoms that have at most one like-element nearest neighbor. Using a randomly mixed, equiatomic HEA NP model, Figure 4 quantifies how the percentage of SA and near-SA sites increases with the number of constituent elements, $N$, for both the overall NP and the surface. For the overall NP, the percentage of strictly SA sites rises from a few percent in ternary systems to approximately half of all atoms in a 15-element HEA. If considering near-SA sites, nearly all atoms are of this type in the case of the 15-element HEA. When considering the surface, which is of key interest for catalytic applications, these percentages increase further (see Supporting Information for details on calculations).

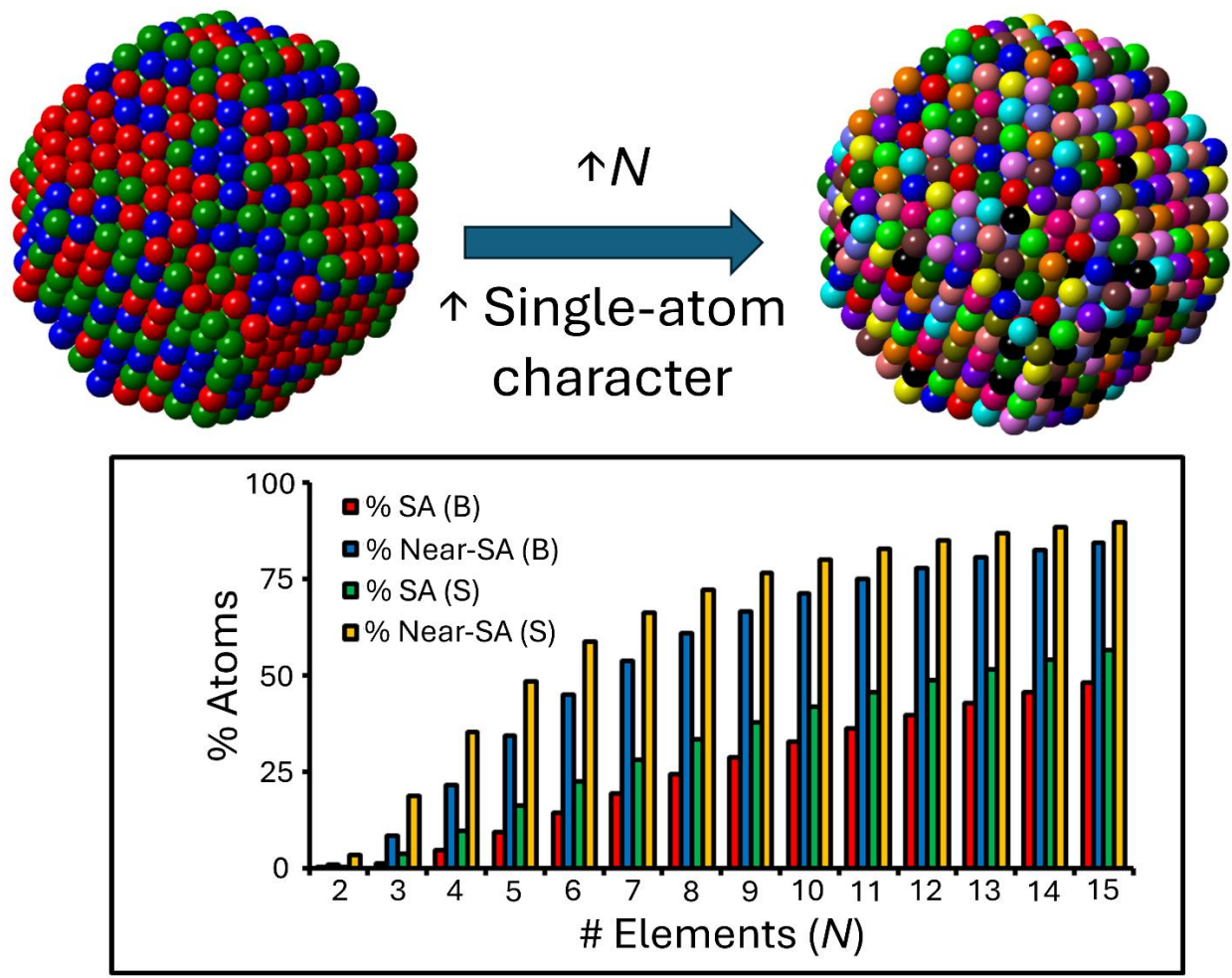

**Figure 4.** Emergence of single-atom and near-single-atom motifs in equiatomic HEA NPs of increasing compositional complexity. Top: representative randomly mixed fcc NPs with $N = 3$ (left) and $N = 15$ (right). Different colors indicate different elements, illustrating the increasing single-atom character as $N$ increases. Bottom: calculated percentages of single-atom (SA) and near-single-atom (near-SA) sites in the bulk (B) and surface (S) atomic sites. SA sites are defined as atoms with no like-element nearest neighbors in the first coordination shell, while near-SA sites have a maximum of one like-element nearest neighbor. As $N$ increases, like-like bonding is naturally suppressed, resulting in a steep increase in both bulk and surface SA fractions. Near-SA sites dominate the surface population in the case of HEAs with high *N*. This trend highlights the remarkable potential of compositionally complex HEAs to serve as high-loading SAA platforms.

Recent studies have begun to explore isolated active sites in HEA-based catalysts, suggesting that the naturally occurring statistical emergence of SA (and near-SA) sites in these systems can be leveraged for unique catalyst design.[115,116] In the longer term, a better understanding of LCO could allow for the manipulation of ordering in the material. This could enable the promotion of SAA-like motifs for catalytically active elements, while the remaining elements act as a host environment to stabilize them. This raises the prospect of designing catalysts that combine the high loading and thermal stability of HEAs with the unique selectivity and electronic structure tuning associated with SAAs.[75,76,117–120] Furthermore, the complex multi-element host structure opens the possibility of synergistic effects that are not possible in conventional binary SAAs.[121] Future work should focus on understanding and controlling LCO to promote the formation of SA and near-SA motifs, in turn enabling the rational design of HEA-based catalysts with optimized activity, selectivity, and durability. Accomplishing this will require robust, element-specific structural characterization capable of resolving LCO in complex multi-element materials.

To achieve this level of structural control, we envision that XAS will become a routine and indispensable component of LCO analysis in HEA research. Its element-specificity, quantitative nature, and compatibility with diverse sample preparation methods make it uniquely suited to disentangle the complexity of modern HEA samples. At present, characterization results reported across the field remain fragmented and largely non-standardized, with different research groups using distinct fitting protocols, metrics, and reporting conventions. Additionally, most studies have utilized ex situ XAS, providing a static picture of a highly dynamic system.

To realize the full potential that XAS offers, the community must work towards broader adoption of XAS for LCO characterization, supported by standardized analysis protocols, reproducible fitting workflows, and open databases of fitting parameters. In the near-term, EXAFS data quality can be improved by utilizing longer k-ranges, simultaneous multi-temperature EXAFS fitting, higher signal-to-noise acquisition, and advanced detection modes such as HERFD. These improvements will increase the precision of calculated CN values and, in turn, strengthen conclusions regarding LCO. In parallel, establishing robust uncertainty-propagation metrics, such as the $\Delta P$ parameter, and encouraging their consistent use across the field will provide a standardized quantitative framework for the comparison of various HEA systems. The continued development and wider adoption of operando and time-resolved XAS will allow the formation and evolution of specific ordering motifs to be directly measured during synthesis and under realistic catalytic operating conditions. Finally, XAS should be combined with complementary characterization techniques and computational modelling to provide the context necessary to interpret XAS results and enable the rational design of HEAs. Together, these efforts will enable standardized, statistically robust LCO measurements in complex multi-element systems. As XAS becomes a standard tool for HEA research, it will provide a common quantitative "language" for describing LCO, thereby accelerating progress towards a more complete understanding of structure–property relationships in these materials.

4. CONCLUSION

In summary, local chemical order (LCO) has emerged as a key structural descriptor for understanding and tuning the properties of HEAs, yet its reliable measurement and interpretation remain challenging. Among the available characterization methods, X-ray absorption spectroscopy (XAS) offers a powerful, element-specific approach for quantifying LCO, and recent advancements involving improved XAS fitting strategies, coordination number-based metrics, and integration with complementary techniques have brought the field closer to a reproducible, quantitative framework. Looking ahead, continued progress will depend on community-wide efforts to standardize analysis protocols, expand operando and time-resolved capabilities, and combine XAS with complementary techniques to overcome its limitations. Advancements in beamline technology, detection schemes, and data analysis pipelines will further enhance data quality and accessibility, positioning XAS to become the routine method of choice for probing LCO. Realizing this vision will transform LCO from a difficult-to-quantify curiosity to a practical design lever, enabling the rational control of atomic configurations and accelerating the discovery of next-generation HEAs for various applications.

## ASSOCIATED CONTENT

**Supporting Information**.

The Supporting Information is available free of charge.

Supporting Information: additional details on calculations involving ΔP values and probability models for estimating single-atom site fractions.

## AUTHOR INFORMATION

**Corresponding Author**

Peng Zhang. E-mail: Peng.Zhang@Dal.ca

**Author Contributions**

The manuscript was written through contributions of all authors. All authors have given approval to the final version of the manuscript.

**Notes**

The authors declare no competing financial interest.

ACKNOWLEDGMENT

P.Z. thanks the NSERC Canada Discovery Grant for funding. CLS@APS facilities (Sector 20-BM) at the Advanced Photon Source (APS) are supported by the U.S. Department of Energy (DOE), NSERC Canada, the University of Washington, the Canadian Light Source (CLS), and the APS. Use of the APS is supported by the DOE under Contract no. DEAC02-06CH11357. The CLS is financially supported by NSERC Canada, CIHR, NRC, and the University of Saskatchewan.

ABBREVIATIONS

APT, atom probe tomography; AET, atomic electron tomography; CN, coordination number; EDS, energy-dispersive X-ray spectroscopy; EELS, electron energy-loss spectroscopy; EXAFS, extended X-ray absorption fine structure; HAADF, high-angle annular dark-field; HEA, high-entropy alloy; HERFD-XANES, high-energy-resolution fluorescence-detected X-ray absorption near-edge structure; LCO, local chemical order; near-SA, near-single-atom; NP, nanoparticle; PDF, pair distribution function; SRO, short-range order; SA, single-atom; SAA, single-atom alloy; STEM, scanning transmission electron microscopy; TEM, transmission electron microscopy; XAS, X-ray absorption spectroscopy; XPS, X-ray photoelectron spectroscopy; XRD, X-ray diffraction.

(102) Martini, A.; Bugaev, A. L.; Guda, S. A.; Guda, A. A.; Priola, E.; Borfecchia, E.; Smolders, S.; Janssens, K.; De Vos, D.; Soldatov, A. V. Revisiting the Extended X-Ray Absorption Fine Structure Fitting Procedure through a Machine Learning-Based Approach. *J. Phys. Chem. A* **2021**, *125*, 7080–7091.

(103) Timoshenko, J.; Lu, D.; Lin, Y.; Frenkel, A. I. Supervised Machine-Learning-Based Determination of Three-Dimensional Structure of Metallic Nanoparticles. *J. Phys. Chem. Lett.* **2017**, *8*, 5091–5098.

(104) Wang, K.; Yu, H.; Lu, X.; Chen, K. Machine Learning-Driven Deconvolution of Mixed X-ray Absorption Spectra. *J. Phys. Chem. A* **2025**, *129*, 7535–7548.

(105) Chill, S. T.; Anderson, R. M.; Yancey, D. F.; Frenkel, A. I.; Crooks, R. M.; Henkelman, G. Probing the Limits of Conventional Extended X-ray Absorption Fine Structure Analysis Using Thiolated Gold Nanoparticles. *ACS Nano* **2015**, *9*, 4036–4042.

(106) Trevorah, R. M.; Chantler, C. T.; Schalken, M. J. Solving self-absorption in fluorescence. *IUCrJ* **2019**, *6*, 586–602.

(107) Joress, H.; Ravel, B.; Anber, E.; Hollenbach, J.; Sur, D.; Hattrick-Simpers, J.; Taheri, M. L.; DeCost, B. Why is EXAFS for complex concentrated alloys so hard? Challenges and opportunities for measuring ordering with X-ray absorption spectroscopy. *Matter* **2023**, *6*, 3763–3781.

(108) Konecny, L.; Vicha, J.; Komorovsky, S.; Ruud, K.; Repisky, M. Accurate X-Ray Absorption Spectra near L- and M-Edges from Relativistic Four-Component Damped Response Time-Dependent Density Functional Theory. *Inorg. Chem.* **2022**, *61*, 830–846.
43

# Supporting Information

# When Atoms Choose Their Neighbors: Element-Specific Views of Local Chemical Order in High-Entropy Alloys

*David Morris[1], Yonggang Yao[2], & Peng Zhang[*,1]*

[1]Department of Chemistry, Dalhousie University, Halifax, NS, Canada.

[2]State Key Laboratory of Materials Processing and Die & Mould Technology, School of Materials Science and Engineering, Huazhong University of Science and Technology, Wuhan, China.

**ΔP Methodology**

This section is derived from our previous work.[1] Briefly, for an element pair A-B, the Warren-Cowley short-range order parameter, $\alpha_{AB}$, is expressed as:[2,3]

$$\alpha_{AB} = 1 - P_{AB}/C_B \qquad (1)$$

where $C_B$ is the molar fraction of element B in the sample, and $P_{AB}$ is the probability of that a nearest neighbor of an A atom is a B atom. Within this sign convention, $\alpha_{AB} > 0$ indicates an enhanced A-B nearest-neighbor association relative to a random solution (chemical ordering), whereas $\alpha_{AB} < 0$ indicates a reduced A-B association relative to random mixing (segregation tendency). Values of $\alpha_{AB} \sim 0$ indicate an elemental distribution consistent with random nearest-neighbor statistics.

In an EXAFS experiment, the fitted coordination numbers represent an ensemble average over all absorber sites of the measured element. Accordingly, $P_{AB}$ is treated as the average fractional occupancy of B within the first coordination shell of absorber A and is obtained from the ratio of the partial and total first-shell coordination numbers:

$$P_{AB} = CN_{AB}/CN_{AM} \qquad (2)$$

where $CN_{AB}$ is the EXAFS-derived first-shell coordination number associated with A-B scattering paths and $CN_{AM}$ is the total first-shell coordination number around absorber A as represented in the fitting model.[2]

While $\alpha_{AB}$ provides a useful formal description of SRO, its interpretation can become less transparent in multicomponent alloys when the goal is to directly compare experimentally derived nearest-neighbor occupancies to a random-alloy baseline. For this purpose, we define a complementary descriptor, ΔP, which measures the deviation of the experimental nearest-neighbor occupancy from that expected for a perfectly mixed (random) alloy of the same composition:

$$\Delta P = (P_{AB} - C_B) * 100\% \qquad (3)$$

where $C_B$ serves as the random-alloy reference because, in an ideal solid solution with no local chemical correlations, the probability of finding B next to A equals the bulk molar fraction of B. Within this definition, $\Delta P > 0$ indicates enrichment of B in the first shell of A relative to random mixing (chemical ordering), whereas $\Delta P < 0$ indicates depletion of B (segregation tendency). The magnitude $|\Delta P|$ provides a convenient measure of the strength of the deviation from random nearest-neighbor statistics.

The ΔP framework is inherently a nearest-neighbor (first-shell metric) because it is derived from first-shell EXAFS coordination numbers. It is therefore most directly applicable when reliable first-shell partial coordination numbers can be obtained for the relevant absorber-neighbor pairs. In cases where chemical correlations extend substantially beyond the first coordination shell (i.e., strong ordering/segregation across multiple coordination shells), explicit multi-shell EXAFS analysis and complementary techniques that probe longer-length-scale correlations will be required to fully capture the extent of ordering. In compositionally complex alloys such as HEAs, practical sensitivity can also be limited by correlations among EXAFS fit parameters, as well as difficulty separating scattering contributions from elements of similar atomic number. Consequently, ΔP values must be interpreted in the context of their associated uncertainties.

Quantitative interpretation of ΔP requires consideration of the uncertainties on the EXAFS-derived coordination numbers. The Artemis software, commonly used for EXAFS fitting across the literature, reports parameter uncertainties that reflect fit quality and parameter correlations, including the number of degrees of freedom and experimental noise.[4] These uncertainties are propagated though Equations 2 and 3 to obtain uncertainty for ΔP, enabling assessment on whether an observed deviation is statistically distinguishable from zero.

**Probability Model for Single-Atom and Near-Single-Atom Site Fractions in Random Alloys**

Figure 4 considers randomly mixed, equiatomic HEA NPs, in which each element has an atomic fraction x = 1/N and all nearest-neighbor identities are assumed to be statistically independent (i.e. the sample is a perfectly mixed random alloy containing no LCO). For a given atomic site, z represents the first-shell CN. For a site occupied by a given element, the probability of a like-element neighbor is x, and the probability of an unlike-element neighbor is 1 – x. If the sample is randomly mixed, then the number of like-element nearest-neighbors follows a binomial distribution.

In this work single-atom (SA) sites are defined as atoms with no like-element neighbors in the first coordination shell. The probability that a site is SA is therefore:

$$P(SA \mid z, N) = (1 - x)^z = (1 - \tfrac{1}{N})^z \qquad (4)$$

In this work near-single-atom (near-SA) sites are defined as atoms with a maximum of one like-element neighbor in the first coordination shell. The probability that a site is near-SA is therefore:

$$P(near - SA \mid z, N) = (1 - x)^z + zx(1 - x)^{z-1} = (1 - \tfrac{1}{N})^z + z(\tfrac{1}{N})(1 - \tfrac{1}{N})^{z-1} \qquad (5)$$

When considering an HEA NP, the first-shell CN will differ significantly between bulk (B) and surface (S) sites, resulting in a higher proportion of SA and near-SA motifs on the surface. In this work, we assume z = 12 for the bulk case, consistent with fcc interior coordination. For the surface case, we use CN values consistent with our previously developed 1289-atom model[1] as an illustrative example rather than a universal description, and weight the probabilities above by the number of surface sites of each CN present in the model, thereby estimating the surface SA and near-SA fractions under random mixing. To reproduce this analysis for a specific material, the CN distribution (particularly for surface sites) should be taken from the system of interest (experimentally or from an appropriate atomistic model).